\documentclass[aps,twocolumn,amsmath,letterpaper]{revtex4}
\usepackage{amssymb}
\usepackage{graphicx}
\usepackage{array}
\usepackage{hhline}
\usepackage{longtable}
\usepackage{array, graphicx, xcolor}

\begin{document}

    \title{Stepwise Positional-Orientational Order and the
    Multicritical-Multistructural Global

     Phase Diagram of the $s=3/2$ Ising Model from Renormalization-Group Theory}

\author{\c{C}a\u{g}{\i}n Yunus$^{1}$, Ba\c{s}ak Renklio\u{g}lu$^{2,3}$, Mustafa Keskin$^4$, and A. Nihat Berker$^{5,6}$}

\affiliation{$^1$Department of Physics, Bo\u{g}azi\c{c}i University,
Bebek 34342, Istanbul, Turkey}

\affiliation{$^2$College of Sciences, Ko\c{c} University, Sar{\i}yer
34450, Istanbul, Turkey}

\affiliation{$^3$Department of Physics, Bilkent University, Bilkent
06533, Ankara, Turkey}

\affiliation{$^4$Department of Physics, Erciyes University, Kayseri
38039, Turkey}

\affiliation{$^5$Faculty of Engineering and Natural Sciences,
Sabanc\i~University, Tuzla 34956, Istanbul, Turkey}

\affiliation{$^6$Department of Physics, Massachusetts Institute of
Technology, Cambridge, Massachusetts 02139, U.S.A.}

\begin{abstract}

The spin-3/2 Ising model, with nearest-neighbor interactions only,
is the prototypical system with two different ordering species, with
concentrations regulated by a chemical potential.  Its global phase
diagram, obtained in $d=3$ by renormalization-group theory in the
Migdal-Kadanoff approximation or equivalently as an exact solution
of a $d=3$ hierarchical lattice, with flows subtended by 40
different fixed points, presents a very rich structure containing
eight different ordered and disordered phases, with more than
fourteen different types of phase diagrams in temperature and
chemical potential. It exhibits phases with orientational and/or
positional order. It also exhibits quintuple phase transition
reentrances. Universality of critical exponents is conserved across
different renormalization-group flow basins, via redundant fixed
points. One of the phase diagrams contains a plastic crystal
sequence, with positional and orientational ordering encountered
consecutively as temperature is lowered.  The global phase diagram
also contains double critical points, first-order and critical lines
between two ordered phases, critical endpoints, usual and unusual
(inverted) bicritical points, tricritical points, multiple
tetracritical points, and zero-temperature criticality and
bicriticality. The 4-state Potts permutation-symmetric subspace is
contained in this model.




\end{abstract}

    \maketitle
    \def\s{\rule{0in}{0.28in}}
    \setlength{\LTcapwidth}{\columnwidth}

\section{Introduction}

\begin{figure*}[]
\centering
\includegraphics*[scale=0.75]{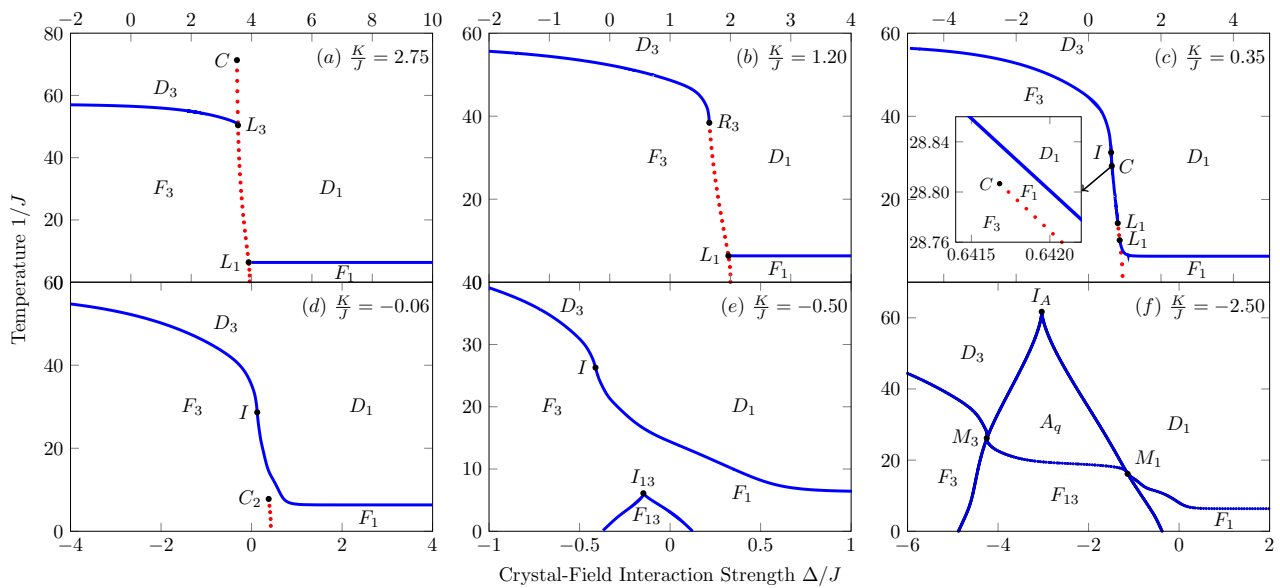}
\caption{(Color online) Temperature versus chemical potential phase
diagrams of the spin-3/2 Ising model for $d=3$, for (a) $K/J =
2.75$, (b) $K/J = 1.20$, (c) $K/J = 0.35$, (d) $K/J = 0.06$, (e)
$K/J = -0.50$, (f) $K/J = -2.50$. The first- and second-order phase
transitions are drawn with dotted and full curves, respectively. The
phases $F_1$ and $F_3$ are ferromagnetically ordered with
predominantly $|s| = 1/2$ and $|s| = 3/2$, respectively.  The phases
$D_1$ and $D_3$ are disordered with predominantly $|s| = 1/2$ and
$|s| = 3/2$, respectively.  The phase $F_{13}$ is positionally and
ferromagnetically ordered and the phase $A_q$ is positionally
ordered and magnetically disordered.  The point $C$ is an ordinary
critical point and the point $C_2$ is a double critical point.  The
points $L_1$ and $L_3$ are critical endpoints. The point $R_3$ is a
tricritical point. The points $M_1$ and $M_3$ are tetracritical
points. The points $I$, $I_{13}$, and $I_A$ each separate two
segments of second-order phase transitions, between the same two
phases, where in spite of renormalization-group flows to different
basins, critical exponent universality is sustained via redundant
fixed points.}
\end{figure*}

The spin-3/2 Ising model, with nearest-neighbor interactions only,
exhibits intricate but physically suggestive phase diagrams, as for
example shown in Fig. 1(f) including three separate ferromagnetic
phases and an only positionally ordered phase, new special points, a
conservancy of the universality principle of critical exponents via
the redundant fixed-point mechanism, and a temperature sequence of
stepwise positional and orientational ordering as in plastic
crystals. Other phase diagram cross-sections of the global phase
diagram, with eight different ordered and disordered phases, include
order-order double critical points, first-order and critical lines
between ordered phases, critical endpoints, usual and unusual
(inverted) bicritical points, tricritical points, different types of
tetracritical points, and zero-temperature criticality and
bicriticality. The permutation-symmetric 4-state Potts subspace is
lodged in this model.

The Hamiltonian of the spin-1/2 Ising model, $-\beta
\mathcal{H}=\Sigma_{\langle ij \rangle} Js_is_j$, where at each site
$i$ there is a spin $s_i=\pm 1$ and the sum is over all pairs of
nearest-neighbor sites, generalizes for the spin-1 Ising model to
\begin{equation}
-\beta \mathcal{H}=\sum_{\langle ij \rangle} [J s_is_j +K s_i^2s_j^2
-\Delta (s_i^2+s_j^2)],
\end{equation}
where $s_i=\pm1, 0$.\cite{BEG} Eq.(1) constitutes the most general
spin-1 Ising model with nearest-neighbor interactions only and no
externally imposed symmetry breaking in the ordering degrees of
freedom.  The global understanding \cite{BerkerWortis,Hoston} of the
phase diagram of the spin-1 Ising model played an important role
through applicability to many physical systems that incorporate
non-ordering degrees of freedom ($s_i=0$) as well as ordering
degrees of freedom ($s_i=\pm1$). The next qualitative step is the
global study of a model that has two different types of local
ordering degrees of freedom, namely the spin-3/2 Ising model
\cite{Sivardiere, Krinsky1, Krinsky2, Bakchich3, Bakchich2,
Izmailian, Benyoussef, Plascak1, Albayrak1, Bakchich4, Bakchich1,
Ozsoy1, Albayrak2, Ekiz, Ozsoy2, Canko1, Plascak2, Canko2, Canko3,
Canko4, Canko5, Keskin1, Keskin2, Keskin3, Keskin4, ElBouziani}:
\begin{multline}
-\beta \mathcal{H}=\sum_{\langle ij \rangle} [(J_1 P_iP_j + J_{13}
(P_iQ_j+Q_iP_j) + J_3 Q_iQ_j) s_is_j\\
+(K_1 P_iP_j + K_{13} (P_iQ_j+Q_iP_j) + K_3 Q_iQ_j) s_i^2s_j^2
-\Delta (s_i^2+s_j^2)],
\end{multline}
where $s_i=\pm 3/2,\pm 1/2$, is the most general spin-3/2 Ising
model with only nearest-neighbor interactions and no externally
imposed symmetry breaking in the ordering degrees of freedom. The
projection operators in Eq.(2) are $P_i = 1 -Q_i = 1 (0)$ for $s_i =
\pm 1/2 (\pm 3/2)$.

Of the models defined above, the spin-1/2 Ising model has a single
critical point on the temperature $J^{-1}$ axis.  The spin-1 Ising
model, in the temperature and chemical potential $\Delta/J$ plane,
has three different types of phase diagrams when the biquadratic
interaction $K$ is non-negative.\cite{BerkerWortis} When negative
biquadratic interactions are considered, nine different types of
phase diagrams are obtained from mean-field theory.\cite{Hoston} We
find in our current work on the spin-3/2 Ising model, using
renormalization-group theory, an extraordinarily rich solution, with
numerous types of phase diagrams in temperature and chemical
potential, exhibiting first- and second-order phase transitions
between eight different variously ordered and disordered phases.

The Hamiltonian of Eq.(2) is expressed as
\begin{equation}
-\beta \mathcal{H}=\Sigma_{\langle ij \rangle} [-\beta
\mathcal{H}(s_i,s_j)],
\end{equation}
where the nearest-neighbor Hamiltonian is
\begin{multline}
-\beta \mathcal{H}(s_i,s_j)=
[J_1 P_iP_j + J_{13}(P_iQ_j+Q_iP_j)\\
+ J_3 Q_iQ_j] s_is_j +[K_1 P_iP_j + K_{13} (P_iQ_j+Q_iP_j)\\
+ K_3 Q_iQ_j] s_i^2s_j^2 -\Delta (s_i^2+s_j^2).
\end{multline}
The transfer matrix, used below, is the exponentiated
nearest-neighbor Hamiltonian
\begin{equation}
T(s_i,s_j) = e^{-\beta \mathcal{H}(s_i,s_j)}.
\end{equation}
We perform the renormalization-group treatment of the system, for
spatial dimension $d$ and length-rescaling factor $b$, using the
Migdal-Kadanoff approximation \cite{Migdal,Kadanoff} or equivalent
exact solution of a $d=3$ hierarchical lattice \cite{BerkerOstlund}.
This calculation is effected by taking the $b^{d-1}$nth power of
each term of the transfer matrix and then by taking the $b$th power
of the resulting matrix.  At each stage, each element of the
resulting matrix is divided by the largest element, which is
equivalent to subtracting an additive constant from the Hamiltonian.
From spin-up-down and nearest-neighbor-interchange symmetries, the
transfer matrix has 6 independent matrix elements, namely
$(T_{33},T_{11},T_{31},T_{1-1},T_{3-1},T_{3-3}$, where the
subscripts refer to the $2s_i,2s_j$ values), one of which is 1 due
to the division mentioned above. Thus, the renormalization-group
flows are in 5-dimensional interaction space. These
renormalization-goup flows are followed until the stable fixed
points of the phases or the unstable fixed points of the phase
transitions are reached, thereby precisely mapping the global phase
diagram from the initial conditions of the variously ending
trajectories.\cite{BerkerWortis} Analysis at the unstable fixed
points yields the order of the phase transitions and the critical
exponents of the second-order transitions. Thus, we have studied the
spin-3/2 Ising model in spatial three dimensions $d=3$ with length
rescaling factor $b=3$, obtaining the global phase diagram, which is
underpinned by 40 renormalization-group fixed points (Table I).
Similar calculations have been done in $d=2$ \cite{Bakchich3,
Bakchich1} and $d=3$ \cite{Bakchich4}.

Our renormalization-group treatment constitutes an exact solution
for hierarchical lattices
\cite{BerkerOstlund,Kaufman1,Kaufman2,McKay,Hinczewski1,BerkerMcKay}
which are being extensively used
\cite{Kaufman,Kotorowicz,Barre,Monthus,Zhang,Shrock,Xu,Hwang2013,Ilker1,Herrmann1,
Herrmann2,Garel,Hartmann,Fortin,Wu,Timonin,Derrida,Thorpe,Efrat,Monthus2,
Hasegawa,Lyra,Singh,Xu2014,Hirose1,Ilker2,Silva,Hotta,Ilker3,Boettcher1,
Demirtas,Boettcher2,Hirose2,Boettcher3,Nandy}.  Our treatment is
simultaneously an approximate solution \cite{Migdal,Kadanoff} for
hypercubic lattices. This approximation for the cubic lattice is an
uncontrolled approximation, as in fact are all renormalization-group
theory calculations in $d=3$ and all mean-field theory calculations.
However, the local summation in position-space technique used here
has been qualitatively, near-quantitatively, and predictively
successful in a large variety of problems, such as arbitrary
spin-$s$ Ising models \cite{BerkerSpinS}, global
Blume-Emery-Griffiths model \cite{BerkerWortis}, first- and
second-order Potts transitions \cite{NienhuisPotts,AndelmanBerker},
antiferromagnetic Potts critical phases
\cite{BerkerKadanoff1,BerkerKadanoff2}, ordering \cite{BerkerPLG}
and superfluidity \cite{BerkerNelson} on surfaces, multiply
reentrant liquid crystal phases \cite{Indekeu,Garland}, chaotic spin
glasses \cite{McKayChaos}, random-field \cite{FalicovRField} and
random-temperature \cite{HuiBerker} magnets, and high-temperature
superconductors \cite{HincsewskiSuperc}.

\section{Global Phase Diagram}

\begin{figure}[]
\centering
\includegraphics*[scale=0.37]{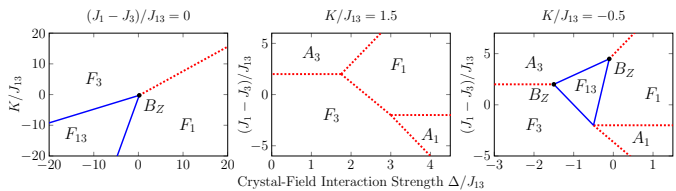}
\caption{(Color online) Zero-temperature ($J, J_{13} \rightarrow
\infty$) phase diagrams. The first- and zero-temperature
second-order phase transitions are drawn with dotted and full
curves, respectively. The phases $F_1$ and $F_3$ are
ferromagnetically ordered with predominantly $|s| = 1/2$ and $|s| =
3/2$, respectively. The phase $F_{13}$ is positionally and
ferromagnetically ordered. The phases $A_1$ and $A_3$ are
antiferromagnetically ordered with predominantly $|s| = 1/2$ and
$|s| = 3/2$, respectively. The points $B_Z$ are zero-temperature
bicritical points which, being at zero temperature, accommodate
boundary lines at finite angles.}
\end{figure}

\begin{figure*}[]
\centering
\includegraphics*[scale=0.75]{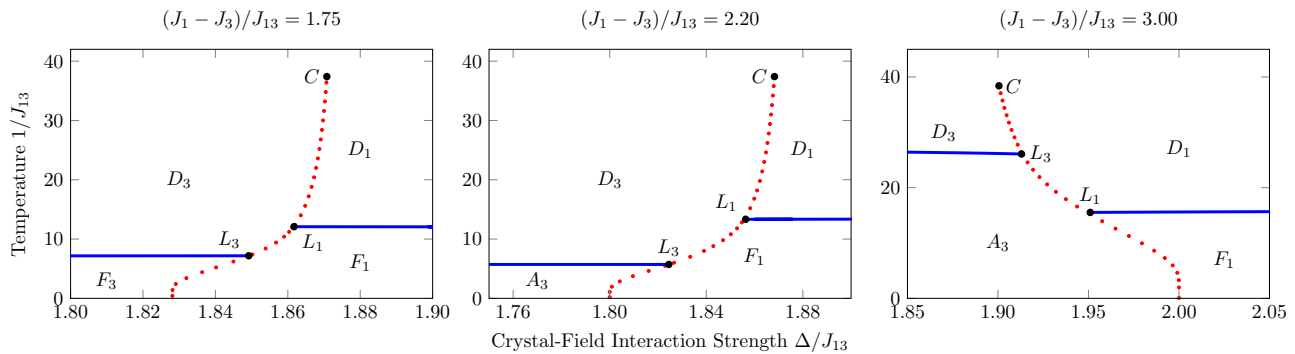}
\caption{(Color online) Phase diagrams evolving from the Fig. 1(a)
topology, with $K/J_{13} = 1.50$.  As $(J_1 - J_3)/J_{13}$ is
increased, the second-order transition temperature to the
ferromagnetic phase $F_3$ falls below the second-order transition
temperature to the ferromagnetic phase $F_1$, as seen for $(J_1 -
J_3)/J_{13} = 1.75$.  Eventually, the ferromagnetic phase $F_3$
disappears at zero temperature and an antiferromagnetic phase $A_3$,
predominantly with $|s_i| = 3/2$, appears from zero temperature, as
seen for $(J_1 - J_3)/J_{13} = 2.20$.  As $(J_1 - J_3)/J_{13}$ is
further increased, the second-order transition temperature to the
antiferromagnetic phase $A_3$ moves above the second-order
transition temperature to the ferromagnetic phase $F_1$, as seen for
$(J_1 - J_3)/J_{13} = 3.00$.}
\end{figure*}

\begin{figure}[]
\centering
\includegraphics*[scale=0.55]{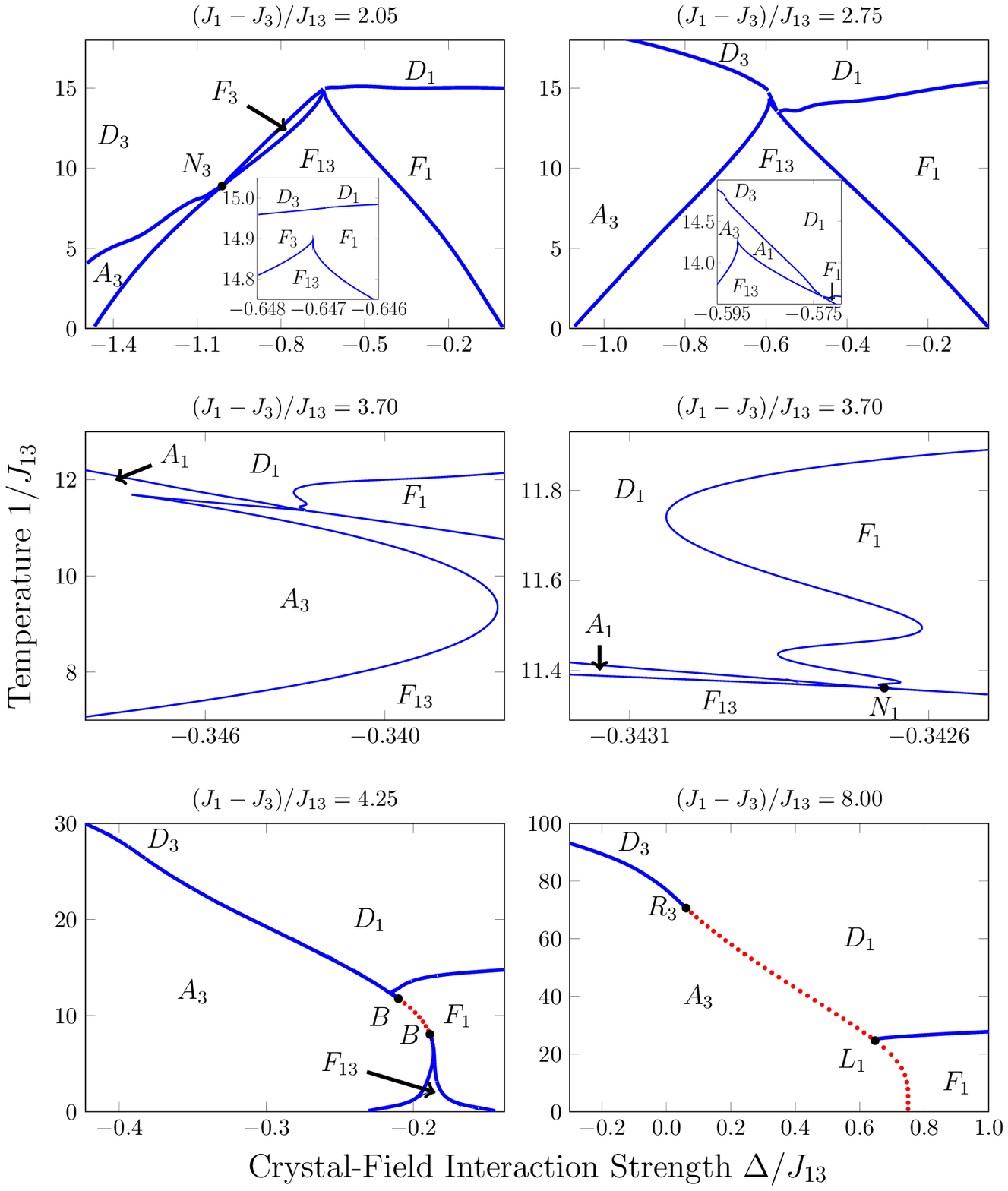}
\caption{(Color online) Phase diagrams evolving from the Fig. 1(e)
topology, with $K/J_{13} = -0.50$.  As $(J_1 - J_3)/J_{13}$ is
increased, the second-order transition temperature to the
ferromagnetic phase $F_3$ is depressed and the antiferromagnetic
phase $A_3$ appears for more negative values of the crystal-field
interaction $\Delta /J_{13}$, where the spin magnitude $|s_i| = 3/2$
is more favored, as seen for $(J_1 - J_3)/J_{13} = 2.05$.  Four
second-order phase transition lines meet at the tetracritical point
$N_3$.  The inset shows that the phase diagram topology near the
maximal temperatures of the phase $F_{13}$ is yet unaltered from
Fig. 1(e). In this phase diagram region, as $(J_1 - J_3)/J_{13}$ is
further increased, the phase $A_3$ replaces $F_3$, which is
eliminated, as seen for $(J_1 - J_3)/J_{13} = 2.75$. The
tetracritical point moves, as $N_1$, to the less negative
crystal-field side of $F_{13}$.  A seen in the phase diagrams for
$(J_1 - J_3)/J_{13} = 3.70$, many phase-transition reentrances occur
in the neighborhood of the tetracritical point $N_1$: As temperature
is lowered, the phase transitions to $F_1$ are quintuply reentrant.
The phase transitions to $F_{13}$ are singly reentrant. It should be
noted that our current calculation, showing these complicated
multiple reentrances, is exact for the hierarchical lattice with
spatial dimension $d=3$. As $(J_1 - J_3)/J_{13}$ is further
increased, the multicritical point $N_1$ splits into usual and
"inverted" bicritical points B connected by a first-order transition
line between the antiferromagnetic phase $A_3$ and the ferromagnetic
phase $F_1$, as seen for $(J_1 - J_3)/J_{13} = 4.75$. As $(J_1 -
J_3)/J_{13}$ is further increased, the higher-temperature bicritical
point splits into a tricritical point $R_3$ and a critical endpoint
$L_1$ and the lower-temperature unusual bicritical point disappears,
along with the phase $F_{13}$, at zero temperature, as seen for
$(J_1 - J_3)/J_{13} = 8.00$.}
\end{figure}

We start by studying $J_1 = J_{13} = J_3 \equiv J$ and $K_1 = K_{13}
= K_3 \equiv K$. Thus, $1/J$ is proportional to temperature and will
be used as our temperature variable. In our system, one of the
ordering species is $|s_i| = 3/2$, the other one is $|s_i| = 1/2$.
The chemical potential $\Delta/J$ (dividing out inverse temperature)
controls the relative amounts of each ordering species. The
biquadratic interaction $K/J$ (again dividing out inverse
temperature) controls the separation/mixing of the two ordering
species. Fig. 1 shows the effects of the biquadratic interaction on
the global phase diagram.

The temperature versus chemical potential phase diagram for large
$K/J$, where separation is favored, is illustrated in Fig. 1(a) with
$K/J = 2.75$. In this phase diagram, two ferromagnetically ordered
phases $F_3$ and $F_1$ are seen at low temperatures, each rich in
one of the ordering species, namely respectively rich in $|s_i| =
3/2$ and $|s_i| = 1/2$. Upon increasing temperature, each
ferromagnetic phase undergoes a second-order phase transition to the
disordered (paramagnetic) phase that is rich in the corresponding
species, respectively $D_3$ and $D_1$. By changing the chemical
potential $\Delta/J$, three different first-order phase transitions
are induced between phases rich in different species: A four-phase
coexistence line between the ferromagnetic phases $F_3$ and $F_1$ at
low temperatures, a three-phase coexistence line between the
ferromagnetic phase $F_3$ and the disordered phase $D_1$ at
intermediate temperatures, and a two-phase coexistence line between
the disordered phases $D_3$ and $D_1$ at high temperatures. Each of
the different first-order fixed points are given in Table I. The
latter first-order transition terminates at high temperature at the
isolated critical point $C$. At intermediate temperatures, both
second-order transition lines terminate on the first-order boundary,
at critical endpoints $L_3$ and $L_1$.  The corresponding hybrid
fixed points, which include both first-order ($y_1=d$) and
second-order ($0<y_2<d$) characteristics \cite{BerkerWortis}, are
given in Table I.

As the biquadratic coupling strength $K/J$ is decreased from large
positive values, lessening the tendency of the two ordering species
to separate, the first-order phase transition line between their
respective disordered phases shrinks, so that the isolated critical
point $C$ and the upper critical endpoint $L_3$ approach each other
and merge, to form the tricritical point $R_3$. The resulting phase
diagram is illustrated in Fig. 1(b) with $K/J = 1.20$.

At lower values of the biquadratic coupling strength $K/J$, a narrow
band of the ferromagnetic $F_1$ phase appears, decoupled from the
main $F_1$ region, between the $F_3$ and $D_1$ regions, as seen in
Fig. 1(c) for $K/J = 0.35$. The first-order phase boundary between
this narrow $F_1$ region and the $F_3$ region extends, at lower
temperature, to the upper critical endpoint $L_1$ and, at higher
temperature, to an isolated critical point $C_2$ as seen in the
inset in Fig. 1(c). This isolated critical point, totally imbedded
in ferromagnetism, is thus a double critical point \cite{Plascak2},
as it mediates between the positively magnetized $F_3$ and $F_1$
and, separately, between the negatively magnetized $F_3$ and $F_1$.
Due to this order-order critical point, it is possible to go
continuously, without encountering a phase transition, between the
ordered $F_3$ and $F_1$ phases. The second-order phase boundary
extending to the upper $L_1$ is composed of to segments, on each
side of the point $I$, separately subtended by the $F_3 - D_3$ and
$F_1 - D_1$ critical fixed points.  The universality principle of
the critical exponents is sustained here by the redundant
\cite{Wegner} fixed-point mechanism: Although these two fixed points
are globally separated in the renormalization-group flow diagram,
they have identical critical exponents (which is furthermore shared
by the fixed point of $I$), as seen in Table I.

At lower values of $K/J$, the two critical endpoints $L_1$ merge and
annihilate.  A single second-order phase boundary, between the $F_3$
or $F_1$ ordered phase at low temperature and the $D_3$ or $D_1$
disordered phase at high temperature, extends across the entire
phase diagram, as seen in Fig. 1(d) for $K/J = -0.06$. The
universality principle is sustained along this second-order phase
boundary by the redundancy of the fixed points, as explained above.
A single first-order boundary forms between the $F_3$ and $F_1$
ordered phases, disconnected from the second-order boundary to the
$D_3$ and $D_1$ disordered phases.  This phase diagram, for $d=3$
from renormalization-group theory, agrees with the phase diagram
previously found for $d=2$ by finite-size scaling and Monte Carlo
\cite{Benyoussef,Plascak1}.

As the biquadratic coupling strength $K/J$ is further decreased,
increasing the tendency of the two ordering species to mix, the
first-order boundary between $F_3$ and $F_1$ shrinks to zero
temperature and thus disappears.  In fact, ordered mixing appears: A
sublattice-wise (i.e., positionally) ordered, as well as
magnetically (i.e., orientationally) ordered ferrimagnetic phase
$F_{13}$ appears at $K/J \leqslant -1/4$.  In this phase, one of two
sublattices is predominantly $|s_i| = 3/2$ and the other sublattice
is predominantly $|s_i| = 1/2$, and the system is magnetized. As
illustrated in Fig. 1(e) for $K/J = -0.50$, the $F_{13}$ phase
occurs at low temperatures and intermediate chemical potentials. The
phase boundary between $F_{13}$ and $F_3$ or $F_1$ is second-order
and remarkably extends to zero-temperature.

For even more negative values of $K/J$, a portion of the
sublattice-ordered phase has erupted through the ferromagnetic
ordering lines and in the process lost ferromagnetic ordering, as
illustrated in Fig. 1(f) with $K/J = -2.50$.  Thus, a new
(antiquadrupolar \cite{Hoston}) phase $A_q$ appears, that is
sublattice-wise (positionally) ordered, but paramagnetic. Each
ordering species predominantly occurs in one of the two sublattices,
with no preferred spin orientation.  This regime offers a phase
diagram topology including four different ordered phases. Two of the
ordered phases are orientationally ordered, one phase is
positionally ordered, and one phase is both orientationally and
positionally ordered. Note that, at intermediate chemical
potentials, as temperature is lowered, the sequence of disordered,
then only positionally ordered, finally positionally and
orientationally ordered phases is encountered, as in plastic crystal
systems.\cite{Pople} Second-order phase transition lines cross at
the tetracritical \cite{Bruce} points $M_3$ and $M_1$. Again, no
violation of universality is seen around the phase $F_{13}$ in Fig.
1(e) or around the phase $A_q$ in Fig. 1(f), the segments on each
side of the points $I_{13}$ and $I_A$ having different fixed points
but same critical exponents (Table I). Phase diagrams similar to
Fig. 1(e,f) have been seen by renormalization-group theory in $d=2,3
 $\cite{Bakchich3,Bakchich1,Bakchich4}. The phase diagram sequence in
Fig.1 (d-f) is in qualitative topological agreement with Bethe
lattice solution \cite{Ekiz}.

The calculated finite-temperature global phase diagram given in Fig.
1 agrees with the zero-temperature phase diagram given in the left
panel of Fig. 2, calculated by ground-state energy crossings. It is
seen that the zero-temperature phase diagram includes a
zero-temperature bicritical point $B_Z$ at $K/J = -1/4, \Delta/J =
3/16$.

\section{Differentiated Species Coupling}

The spin-3/2 Ising model carries an even richer structure of phase
diagrams, accessed by differentiating the interaction constants in
Eq. (2). We give here two sequences of phase diagrams with
$J_1>J_3$, $J_{13} = (J_1+J_3)/2$.

\subsection{Phase Diagrams Evolving from\\
the Figure 1(a) Topology}

Figure 3 shows phase diagrams with $K/J_{13} = 1.50$.  As $(J_1 -
J_3)/J_{13}$ is increased, the second-order transition temperature
to the ferromagnetic phase $F_3$ falls below the second-order
transition temperature to the ferromagnetic phase $F_1$, as seen for
$(J_1 - J_3)/J_{13} = 1.75$.  Eventually, the ferromagnetic phase
$F_3$ disappears at zero temperature and an antiferromagnetic phase
$A_3$, predominantly with $|s_i| = 3/2$, appears from zero
temperature, as seen for $(J_1 - J_3)/J_{13} = 2.20$.  As $(J_1 -
J_3)/J_{13}$ is further increased, the second-order transition
temperature to the antiferromagnetic phase $A_3$ moves above the
second-order transition temperature to the ferromagnetic phase
$F_1$, as seen for $(J_1 - J_3)/J_{13} = 3.00$.

The finite-temperature phase diagrams of Fig. 3 are consistent with
the corresponding zero-temperature phase diagram, in the middle
panel of Fig. 2. Conversely and not shown here, when $(J_1 -
J_3)/J_{13}$ is made negative, an antiferromagnetic phase $A_1$,
predominantly with $|s_i| = 1/2$, similarly appears, as seen in Fig.
2.

\subsection{Phase Diagrams Evolving from\\
the Figure 1(e) Topology}

Figure 4 shows phase diagrams with $K/J_{13} = -0.50$.  As $(J_1 -
J_3)/J_{13}$ is increased, the second-order transition temperature
to the ferromagnetic phase $F_3$ is depressed and the
antiferromagnetic phase $A_3$ appears for more negative values of
the crystal-field interaction $\Delta /J_{13}$, where the spin
magnitude $|s_i| = 3/2$ is more favored, as seen for $(J_1 -
J_3)/J_{13} = 2.05$.  Four second-order phase transition lines meet
at the tetracritical point $N_3$.  The inset shows that the phase
diagram topology is unaltered near the maximal temperatures of the
phase $F_{13}$. In this phase diagram region, as $(J_1 -
J_3)/J_{13}$ is further increased, $A_3$ replaces $F_3$, which is
eliminated, as seen for $(J_1 - J_3)/J_{13} = 2.75$.  The
tetracritical point moves, as $N_1$, to the less negative
crystal-field side of $F_{13}$.  As seen in the phase diagrams for
$(J_1 - J_3)/J_{13} = 3.70$, many phase transition reentrances
\cite{Cladis,Hardouin,Indekeu,Garland,Netz,Kumari,Caflisch} occur in
the neighborhood of the tetracritical point $N_1$: As temperature is
lowered, the phase transitions to $F_1$ are quintuply reentrant. The
phase transitions to $F_{13}$ are singly reentrant. Previously, up
to quadruply reentrant phase transitions have been found for liquid
crystal systems \cite{Cladis,Hardouin,Indekeu,Garland,Netz,Kumari}
and surface systems \cite{Caflisch}. Much higher reentrances have
been calculated in the high-$T_C$ superconductivity tJ
model.\cite{Falicov}. It should be noted that our current
calculation, showing these complicated reentrances, is exact for the
hierarchical lattice with spatial dimension $d=3$. As $(J_1 -
J_3)/J_{13}$ is further increased, the tetracritical point $N_1$
splits into two bicritical points B connected by a first-order
transition line between the antiferromagnetic phase $A_3$ and the
ferromagnetic phase $F_1$, as seen in Fig. 4 for $(J_1 - J_3)/J_{13}
= 4.75$. At a (non-zero-temperature) bicritical point, normally, two
high-temperature second-order boundaries and one low-temperature
first-order boundary meet tangentially.\cite{Bruce} In our present
phase diagram, the opposite temperature ordering occurs at the
lower-temperature bicritical point. Thus, this is an "inverted
bicritical point". As $(J_1 - J_3)/J_{13}$ is further increased, the
higher-temperature bicritical point splits into a tricritical point
$R_3$ and a critical endpoint $L_1$ and the lower-temperature
inverted bicritical point disappears, along with the phase $F_{13}$,
at zero temperature, as seen in Fig. 4 for $(J_1 - J_3)/J_{13} =
8.00$.

Thus, both multicritical points, bicritical and tetracritical, of
the classic coupled-order-parameter problem \cite{Bruce} is
contained within the spin-3/2 Ising model. The finite-temperature
phase diagrams of Fig. 4 are consistent with the corresponding
zero-temperature phase diagram, in the right panel of Fig. 2.

\begin{figure}[]
\centering
\includegraphics*[scale=1]{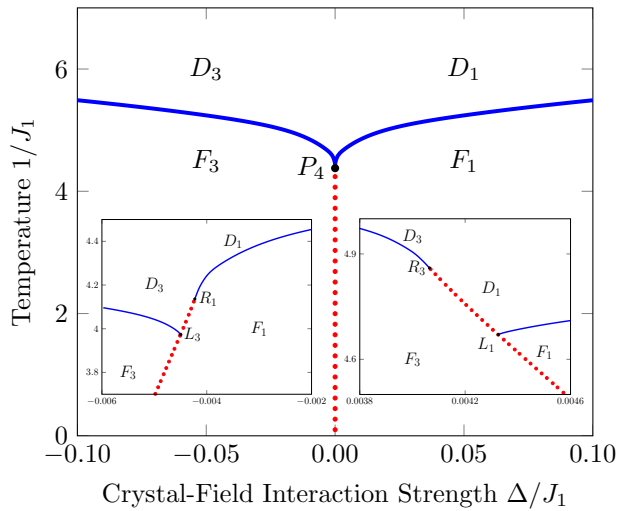}
\caption{(Color online) The phase diagram for the 4-state Potts side
conditions $4 J_1 = 36 J_3 = K_1 = 81 K_3, J_{13} = K_{13} = 0$. In
this figure, $\Delta /J_1 = 0$ is the 4-state Potts subspace, where
the system is permutation symmetric with respect to the 4 states
$s_i=\pm 3/2,\pm1/2$. Varying $\Delta /J_1$ from zero gives the
symmetric bicritical phase diagram around the Potts transition point
$P_4$. This phase diagram is exact for the $d=3$ hierarchical
lattice, but approximate for the cubic lattice, as explained in the
text. Away from the 4-Potts condition, e.g. for $3.5 J_1 = 36 J_3 =
K_1 = 81 K_3$ and $4.5 J_1 = 36 J_3 = K_1 = 81 K_3$, shown in the
left and right insets respectively, the symmetric bicritical point
$P_4$ is replaced, respectively, by tricritical points $R_3$ and
$R_1$ and critical endpoints $L_1$ and $L_3$ in asymmetric phase
diagrams.}
\end{figure}

\section{4-State Potts Subspace}

Returning to the most general spin-3/2 Ising Hamiltonian in Eq.(2),
for
\begin{equation}
4 J_1 = 36 J_3 = K_1 = 81 K_3, \quad J_{13} = K_{13} = \Delta = 0,
\end{equation}
the model reduces to the 4-state Potts model, with Hamiltonian
\begin{equation}
 -\beta \mathcal{H}=  \frac{J_1}{2} \sum_{\langle ij \rangle} \delta
 (s_i,s_j),
\end{equation}
where the Kronecker delta function is $\delta(s_i,s_j) = 1 (0)$ for
$s_i = s_j (s_i \neq s_j)$.

Figure 5 gives the calculated phase diagram in temperature $1/J_1$
and chemical potential $\Delta/J_1$, for the 4-state Potts side
conditions $4 J_1 = 36 J_3 = K_1 = 81 K_3, J_{13} = K_{13} = 0$. In
this figure, $\Delta /J_1 = 0$ is the 4-state Potts subspace, where
the system is permutation symmetric with respect to the 4 states
$s_i=\pm 3/2,\pm1/2$.  Varying $\Delta /J_1$ from zero gives the
"symmetric bicritical" phase diagram around the Potts multicritical
point. This phase diagram is exact for the $d=3$ hierarchical
lattice, but approximate for the cubic lattice.  For the cubic
lattice, from 3-state Potts model analogy \cite{StraleyFisher}, we
expect short first-order segments on each phase boundary leading to
the 4-state Potts transition, which occurs as first-order with
5-phase coexistence. In renormalization-group theory, this
first-order behavior is revealed by the accounting of local disorder
as effective vacancies \cite{NienhuisPotts,AndelmanBerker}.  For the
spin-3/2 Ising model, the resulting renormalization-group flows
would be in the space of the spin-2 Ising model. In Fig. 5, for $3.5
J_1 = 36 J_3 = K_1 = 81 K_3$ and $4.5 J_1 = 36 J_3 = K_1 = 81 K_3$,
shown in the left and right insets respectively, the symmetric Potts
transition point $P_4$ is replaced, respectively, by tricritical
points $R_3$ and $R_1$ and critical endpoints $L_1$ and $L_3$ in
asymmetric phase diagrams.

\section {Conclusion}

We have obtained the full unified global phase diagram of the
spin-3/2 Ising model in $d=3$, deriving extremely rich structures
and fully showing the logical continuity among these complicated
structures. Thus, renormalization-group theory reveals eight
different orientationally and/or positionally ordered and disordered
phases; first- and second-order phase transitions; double critical
points, first-order and critical lines between ordered phases,
critical endpoints, usual and unexpected (inverted) bicritical
points, tricritical points, different tetracritical points, very
high (quintuple) phase boundary reentrances, and zero-temperature
criticality and bicriticality. Fourteen different phase diagram
topologies, in the temperature and chemical potential variables, are
presented here. The renormalization-group flows yielding this
multicritical, multistructural global phase diagram are governed by
40 different fixed points (Table I). Globally distant in flow space,
redundant fixed points nevertheless conserve the universality of
critical exponents. The imbedding of the 4-state Potts symmetric
subspace and phase transition is made explicit.

\begin{acknowledgments}
We thank Tolga \c{C}a\u{g}lar for his help. Support by the Alexander
von Humboldt Foundation, the Scientific and Technological Research
Council of Turkey (T\"UBITAK), and the Academy of Sciences of Turkey
(T\"UBA) is gratefully acknowledged.
\end{acknowledgments}

\begin{table*}[h!]

I. Stable Fixed Points: Phase Sinks\\
\begin{tabular}{|c|c|c|c|}
\hline
$F_3$ & $F_1$ & $A_3$ & $A_1$ \\
Long Ferro & Short Ferro & Long Antiferro & Short Antiferro\\
$\left( \begin{array}{cccc} 1 & 0 & 0 & 0 \\ 0 & 0 & 0 & 0 \\
0 & 0 & 0 & 0 \\ 0 & 0 & 0 & 1 \end{array} \right)$ & $\left(
\begin{array}{cccc} 0 & 0 & 0 & 0 \\ 0 & 1 & 0 & 0 \\ 0 & 0 & 1 & 0
\\ 0 & 0 & 0 & 0 \end{array} \right)$ & $\left( \begin{array}{cccc} 0 & 0 & 0 & 1 \\ 0 & 0 & 0 & 0 \\
0 & 0 & 0 & 0 \\ 1 & 0 & 0 & 0 \end{array} \right)$ & $\left( \begin{array}{cccc} 0 & 0 & 0 & 0 \\ 0 & 0 & 1 & 0 \\
0 & 1 & 0 & 0 \\ 0 & 0 & 0 & 0 \end{array} \right)$\\
\hline

$F_{13}$ & $A_q$ & $D_3$ & $D_1$ \\
Mixed Ferro & Plastic Crystal & Long Disordered & Short Disordered \\
$\left( \begin{array}{cccc} 0 & 1 & 0 & 0 \\ 1 & 0 & 0 & 0 \\ 0 & 0
& 0 & 1
\\ 0 & 0 & 1 & 0 \end{array}\right)$ & $\left( \begin{array}{cccc} 0 & 1 & 1 & 0 \\ 1
& 0 & 0 & 1 \\ 1 & 0 & 0 & 1 \\ 0 & 1 & 1 & 0 \end{array} \right)$ &
$\left( \begin{array}{cccc} 1 & 0 & 0 & 1 \\ 0 & 0 & 0 & 0 \\ 0 & 0
& 0 & 0 \\ 1 & 0 & 0 & 1 \end{array} \right)$
& $\left( \begin{array}{cccc} 0 & 0 & 0 & 0 \\ 0 & 1 & 1 & 0 \\  0 & 1 & 1 & 0 \\ 0 & 0 & 0 & 0 \end{array} \right) $\\
\hline

\end{tabular}

\quad\\
II. Singly Unstable Fixed Points: Attractors of Second-Order Phase Boundaries and their Relevant Exponent $y_T$\\
\begin{tabular}{|c|c|c|c|c|}
        \hline
        $F_3 - D_3$ & $F_1 - D_1$ & $F_{13} - A_q$ & $A_3 - D_3$ & $A_1 - D_1$ \\
        $y_T = 0.9260$ & $y_T = 0.9260$ & $y_T = 0.9260$ & $y_T = 0.9260$ & $y_T = 0.9260$\\

        $\left( \begin{array}{cccc} 1 & 0 & 0 & t \\ 0 & 0 & 0 & 0 \\ 0 & 0 & 0 & 0 \\ t & 0 & 0 & 1 \end{array} \right)$

        & $\left( \begin{array}{cccc} 0 & 0 & 0 & 0 \\ 0  & 1 & t & 0\\ 0 & t & 1 & 0 \\ 0 & 0 & 0 & 0 \end{array} \right)$

&$\left( \begin{array}{cccc} 0 & 1 & t & 0 \\ 1 & 0 & 0 & t \\ t &
0& 0 & 1 \\ 0 & t & 1 & 0 \end{array} \right)$

& $\left( \begin{array}{cccc} t & 0 & 0 & 1 \\ 0 & 0 & 0 & 0 \\ 0 &
0 & 0 & 0 \\ 1 & 0 & 0 & t \end{array} \right)$

        & $\left( \begin{array}{cccc} 0 & 0 & 0 & 0 \\ 0 & t & 1 & 0 \\ 0 & 1 & t & 0 \\ 0 & 0 & 0 & 0 \end{array} \right)$\\
        \hline

        $F_3 - F_{13}$ & $F_{13} - F_1$ & $D_3 - A_q$ & $A_q - D_1$ &\\
        $y_T = 1.8104$ & $y_T = 1.8104$ & $y_T = 1.8104$ & $y_T = 1.8104$ & \\

        $\left( \begin{array}{cccc} v & 1 & 0 & 0 \\ 1 & w & 0 & 0 \\ 0 & 0 & w & 1 \\ 0 & 0 & 1 & v \end{array} \right)$
        & $\left( \begin{array}{cccc} w & 1 & 0 & 0 \\ 1 & v & 0 & 0 \\ 0 & 0 & v & 1 \\ 0 & 0 & 1 & w \end{array} \right)$
        & $\left( \begin{array}{cccc} v & 1 & 1 & v \\ 1 & w & w & 1 \\ 1 & w & w & 1 \\ v & 1 & 1 & v \end{array} \right)$
        & $\left( \begin{array}{cccc} w & 1 & 1 & w \\ 1 & v & v & 1 \\ 1 & v & v & 1 \\ w & 1 & 1 & w \end{array} \right)$ & \\
        \hline
        \end{tabular}

\quad\\
III. Singly Unstable Fixed Points: Attractors of First-Order Phase Boundaries with Relevant Exponent $y_T = d$\\
\begin{tabular}{|c|c|c|c|}
        \hline
        $F_3 - F_1$ & $D_3 -D_1$ & $F_3 -D_1$ & $F_1 - D_3$ \\
        $y_T = d$ & $y_T = d$ & $y_T = d$ & $y_T = d$ \\
        $\left( \begin{array}{cccc} 1 & 0 & 0 & 0 \\ 0 & 1 & 0 & 0 \\ 0 & 0 & 1 & 0 \\ 0 & 0 & 0 & 1 \end{array} \right)$
        & $\left( \begin{array}{cccc} 1 & 0 & 0 & 1 \\ 0 & 1 & 1 & 0 \\ 0 & 1 & 1 & 0 \\ 1 & 0 & 0 & 1 \end{array} \right)$
        & $\left( \begin{array}{cccc} 1 & 0 & 0 & 0 \\ 0 & u & u & 0 \\ 0 & u & u & 0 \\ 0 & 0 & 0 & 1 \end{array} \right)$
        & $\left( \begin{array}{cccc} u & 0 & 0 & u \\ 0 & 1 & 0 & 0 \\ 0 & 0 & 1 & 0 \\ u & 0 & 0 & u \end{array} \right)$\\
        \hline
        \end{tabular}
\quad\\
IV. Singly Unstable Fixed Points: Attractors of Smooth Continuation (Null) Lines\\
\begin{tabular}{|c|c|}
        \hline
        $F_3 - F_1$ & $D_3 -D_1$ \\
        $\left( \begin{array}{cccc} 1 & 1 & 0 & 0 \\ 1 & 1 & 0 & 0 \\ 0 & 0 & 1 & 1 \\ 0 & 0 & 1 & 1 \end{array} \right)$
        & $\left( \begin{array}{cccc} 1 & 1 & 1 & 1 \\ 1 & 1 & 1 & 1 \\ 1 & 1 & 1 & 1 \\ 1 & 1 & 1 & 1 \end{array} \right)$\\
        \hline
        \end{tabular}

\quad\\
V. Multiply Unstable Fixed Points: Attractors of Multicritical Points and their Leading 2 Relevant Exponents $y_{T1}, y_{T2}$ \\
\begin{tabular}{|c|c|c|c|c|c|}
\hline
$L_1$ Critical Endpoint & $L_3$ Critical Endpoint & $B$ Bicritical \\
$y_{T1} = d$, $y_{T2} = 0.9260$ & $y_{T1} = d$, $y_{T2} = 0.9260$ & $y_{T1} = 2.4649$, $y_{T2} = 1.0000$ \\

$\left( \begin{array}{cccc} 1.0212 & 0 & 0 & 0 \\ 0 & 1 & t & 0 \\ 0
& t & 1 & 0 \\ 0 & 0 & 0 & 1.0212 \end{array} \right)$

& $\left( \begin{array}{cccc} 1 & 0 & 0 & t \\ 0  & 1.0212 & 0 & 0\\
0 & 0 & 1.0212 & 0 \\ t & 0 & 0 & 1 \end{array} \right)$

&$\left( \begin{array}{cccc} 0.3347 & 1 & 0.3347 & 1 \\ 1 & 1 & 0.3347 & 0.3347 \\
0.3347 & 0.3347 & 1 & 1 \\ 1 & 0.3347 & 1 & 0.3347 \end{array} \right)$\\
\hline

$R_1$ Tricritical & $R_3$ Tricritical & $P_4$ 4-Potts\\
$y_{T1} = 2.1733$, $y_{T2} = 0.7352$ & $y_{T1} = 2.1733$, $y_{T2} = 0.7352$ & $y_{T1} = 2.5434$, $y_{T2} = 1.0667$\\

$\left( \begin{array}{cccc} 0.9474 & 0.9316 & 0.9316 & 0.9474 \\ 0.9316 & 1 & 0.8344 & 0.9316 \\
0.9316 & 0.8344 & 1 & 0.9316 \\ 0.9474 & 0.9316 & 0.9316 & 0.9474
\end{array} \right)$

& $\left( \begin{array}{cccc} 1 & 0.9316 & 0.9316 & 0.8344 \\ 0.9316 & 0.9474 & 0.9474 & 0.9316 \\
0.9316 & 0.9474 & 0.9474 & 0.9316 \\ 0.8344 & 0.9316 & 0.9316 & 1
\end{array} \right)$

& $\left( \begin{array}{cccc} 1 & 0.8926 & 0.8926 & 0.8926 \\ 0.8926 & 1 & 0.8926 & 0.8926 \\
0.8926 &
0.8926 & 1 & 0.8926 \\ 0.8926 & 0.8926 & 0.8926 & 1 \end{array} \right)$\\

\hline

$M_1$ Tetracritical & $M_3$ Tetracritical & $N_1$ Tetracritical\\
$y_{T1} = 1.8104$, $y_{T2} = 0.9260$ & $y_{T1} = 1.8104$, $y_{T2} = 0.9260$ & $y_{T1} = 2.0000$, $y_{T2} = 1.6805$ \\

$\left( \begin{array}{cccc} 0.4099 & 1 & 0.9243 & 0.3789 \\ 1 & 0.9300 & 0.8596 & 0.9243 \\
0.9243 & 0.8596 & 0.9300 & 1 \\ 0.3789 & 0.9243 & 1 & 0.4099
\end{array} \right)$

& $\left( \begin{array}{cccc} 0.9300 & 1 & 0.9243 & 0.8596 \\ 1  & 0.4099 & 0.3789 & 0.9243\\
0.9243 & 0.3789 & 0.4099 & 1 \\ 0.8596 & 0.9243 & 1 & 0.9300
\end{array} \right)$

& $\left( \begin{array}{cccc} 0.3228 & 1 & 0.2960 & 0.3228 \\ 1 & 0.9170 & 0.9170 & 0.2960 \\
0.2960 & 0.9170 & 0.9170 & 1 \\ 0.3228 & 0.2960 & 1 & 0.3288 \end{array} \right)$\\

\hline

$N_3$ Tetracritical & I Interceding & $I_{13}$ and $I_A$ Interceding\\
$y_{T1} = 2.0000 $, $y_{T2} = 1.6805$ & $y_{T1} = 2.5732$, $y_{T2} = 0.9260$ & $y_{T1} = 1.4268$, $y_{T2} = 0.9260$ \\

$\left( \begin{array}{cccc} 0.9170 & 1 & 0.2960 & 0.9170 \\ 1 & 0.3228 & 0.3228 & 0.2960 \\
0.2960 & 0.3228 & 0.3228 & 1 \\ 0.9170 & 0.2960 & 1 & 0.9170
\end{array} \right)$

& $\left( \begin{array}{cccc} 1 & 0.8543 & t & t \\ 0.8543 & 1 & t & t \\
t & t & 1 & 0.8543 \\ t & t & 0.8543 & 1
\end{array} \right)$

&$\left( \begin{array}{cccc} t & 1 & 0 & 0 \\ 1 & t & 0 & 0 \\ 0 & 0
& t & 1 \\ 0 & 0 & 1 & t \end{array} \right)$ and
 $\left( \begin{array}{cccc} t & 1 & 1 & t \\ 1 & t & t & 1 \\ 1 & t & t & 1 \\ t & 1 & 1 & t \end{array} \right)$\\

\hline

\end{tabular}

\caption{Fixed points underpinning the renormalization-group flows
determining the global phase diagram of the $s=3/2$ Ising model. In
this Table, the matrix elements are $t=0.9243, u=0.9481, v=0.9300,
w=0.4099$. The fixed point for the isolated critical point in Fig.
1(a) has thermal exponent $y_T = 0.9260$ along the first-order
transition direction and magnetic exponent $y_H = 2.5732$ orthogonal
to the first-order transition direction. The fixed point for the
isolated double critical points in Figs. 1(c,d) has thermal exponent
$y_T = 0.9260$ along the first-order transition direction and
magnetic exponent $y_H = 2.5732$ orthogonal to the first-order
transition direction, again realizing critical exponent universality
via redundant renormalization-group fixed points. Not shown are the
critical endpoint fixed points for the antiferromagnetic transitions
of the two species.}

\end{table*}

\end{document}